\documentclass[twocolumn, amsmath, amssymb, aps, prx, superscriptaddress, floatfix, 10pt]{revtex4-2}

\usepackage{microtype}
\usepackage[english]{babel}
\usepackage{float}
\usepackage{wrapfig}
\usepackage{graphicx}% Include figure files
\usepackage{dcolumn}% Align table columns on decimal point
\usepackage{bm}% bold math
\usepackage{hyperref}% add hypertext capabilities
\hypersetup{
    colorlinks=true,
    linkcolor=blue,
    citecolor=blue,
    urlcolor=blue 
}

\begin{document}

\preprint{APS/123-QED}
\date{\today}

\title{Highly-efficient electron ponderomotive acceleration in underdense plasmas}

\author{L.~Martelli}
\email{lorenzo.martelli295@gmail.com}
\affiliation{LOA, \'Ecole Polytechnique, ENSTA ParisTech, CNRS, Universit\'e Paris-Saclay, 828 Boulevard des Mar\'echaux, 91762 Palaiseau Cedex, France}
\affiliation{THALES AVS - MIS, 2 Rue Marcel Dassault, Vélizy-Villacoublay, 78140, France}%
\author{O.~Kononenko}
\affiliation{LOA, \'Ecole Polytechnique, ENSTA ParisTech, CNRS, Universit\'e Paris-Saclay, 828 Boulevard des Mar\'echaux, 91762 Palaiseau Cedex, France}
\author{I.~A.~Andriyash}
\affiliation{LOA, \'Ecole Polytechnique, ENSTA ParisTech, CNRS, Universit\'e Paris-Saclay, 828 Boulevard des Mar\'echaux, 91762 Palaiseau Cedex, France}
\author{J.~Wheeler}
\affiliation{LOA, \'Ecole Polytechnique, ENSTA ParisTech, CNRS, Universit\'e Paris-Saclay, 828 Boulevard des Mar\'echaux, 91762 Palaiseau Cedex, France}
\author{J.~Gautier}
\affiliation{LOA, \'Ecole Polytechnique, ENSTA ParisTech, CNRS, Universit\'e Paris-Saclay, 828 Boulevard des Mar\'echaux, 91762 Palaiseau Cedex, France}
\author{J.-P~Goddet}
\affiliation{LOA, \'Ecole Polytechnique, ENSTA ParisTech, CNRS, Universit\'e Paris-Saclay, 828 Boulevard des Mar\'echaux, 91762 Palaiseau Cedex, France}
\author{A.~Tafzi}
\affiliation{LOA, \'Ecole Polytechnique, ENSTA ParisTech, CNRS, Universit\'e Paris-Saclay, 828 Boulevard des Mar\'echaux, 91762 Palaiseau Cedex, France}
\author{R.~Lahaye}
\affiliation{LOA, \'Ecole Polytechnique, ENSTA ParisTech, CNRS, Universit\'e Paris-Saclay, 828 Boulevard des Mar\'echaux, 91762 Palaiseau Cedex, France}
\author{C.~Giaccaglia}
\affiliation{LOA, \'Ecole Polytechnique, ENSTA ParisTech, CNRS, Universit\'e Paris-Saclay, 828 Boulevard des Mar\'echaux, 91762 Palaiseau Cedex, France}
\affiliation{Institut Curie, Inserm U 1021-CNRS UMR 3347, University Paris-Saclay, PSL Research University, Centre Universitaire, Orsay, 91471, France}%
\author{A.~Flacco}
\affiliation{LOA, \'Ecole Polytechnique, ENSTA ParisTech, CNRS, Universit\'e Paris-Saclay, 828 Boulevard des Mar\'echaux, 91762 Palaiseau Cedex, France}
\author{V.~Tomkus}
\author{M.~Mackevičiūtė}
\author{J.~Dudutis}
\author{V.~Stankevic}
\author{P.~Gečys}
\author{G.~Račiukaitis}
\affiliation{FTMC - Center for Physical Sciences and Technology, Savanoriu Ave. 231, Vilnius, LT-2300, Lithuania}
\author{H.~Kraft}
\author{X.~Q.~Dinh}
\affiliation{THALES AVS - MIS, 2 Rue Marcel Dassault, Vélizy-Villacoublay, 78140, France}%
\author{C.~Thaury}
\email{cedric.thaury@ensta-paris.fr}
\affiliation{LOA, \'Ecole Polytechnique, ENSTA ParisTech, CNRS, Universit\'e Paris-Saclay, 828 Boulevard des Mar\'echaux, 91762 Palaiseau Cedex, France}

\begin{abstract}
Laser-plasma accelerators represent a promising technology for future compact accelerating systems, enabling the acceleration of tens of pC to above $1\,$GeV over just a few centimeters. Nonetheless, these devices currently lack the stability, beam quality and average current of conventional systems. While many efforts have focused on improving acceleration stability and quality, little progress has been made in increasing the beam's average current, which is essential for future laser-plasma-based applications. In this paper, we investigate a laser-plasma acceleration regime aimed at increasing the beam average current with energies up to few-MeVs, efficiently enhancing the beam charge. We present experimental results on configurations that allow reaching charges of $5-30\,$nC and a maximum conversion efficiency of around $14\,$\%. Through comprehensive Particle-In-Cell simulations, we interpret the experimental results and present a detailed study on electron dynamics. From our analysis, we show that most electrons are not trapped in a plasma wave; rather, they experience ponderomotive acceleration. Thus, we prove the laser pulse as the main driver of the particles' energy gain process.
\end{abstract}

\maketitle

\section{\label{sec:level1}Introduction}
Since their proposal~\cite{Tajima}, laser-plasma accelerators (LPA) have interested the scientific community for their ability to produce accelerating gradients $10^3 - 10^4$ times those of conventional systems (i.e., $\sim100\,$MV/m). The extremely intense electric field would allow for a scaling-down of future accelerators, representing a cost-effective alternative to state-of-the-art linear accelerators (LINACs) and radio frequency cavities. The production of stable, low-emittance, and highly energetic monochromatic LPA electron beams~\cite{Faure2006, Leemans2006, Gonsalves2019, Oubrerie2022}, as well as recent optimization studies~\cite{Maier2020, Jalas2021}, have dominated the literature. While these results demonstrate the potential of laser-plasma accelerators, their beam properties are still far from those of conventional systems. For instance, current medical electron LINACs can produce up to $10-10^2\,\mu$A, while LPAs deliver only tens-nA~\cite{Couperus, Rovige2020}. Naturally, one direct method of increasing the average current of LPAs comes from increasing the beam's charge. Using high-Z gases such as nitrogen and argon is proven as an effective way of enhancing the charge up to the nC-level~\cite{EGuillaume, JGotzfried}. Recently, using a $27\,$TW, Ti:Sapphire laser and a pure nitrogen plasma with a density $n_e\approx3.6\times10^{19}\,$cm$^{-3}$, it was possible to accelerate around $15\,$nC to few-MeVs, with divergences exceeding $100$s\,mrad~\cite{JFeng}. This potentially paves the way for $\mu$A-level LPAs, marking an important milestone for laser-plasma-based applications. The acceleration of such beams is believed to be associated with the ionization injection of electrons in multiple plasma periods. Once trapped, the particles are subsequently accelerated by the plasma electric field, producing large energy spectra up to a few-tens of MeV.

In this work, we explore the interaction between a superintense laser pulse and a pure nitrogen plasma with densities in excess of $n_e=0.01\,n_c$, where $n_c(\text{cm$^{-3}$}) = 1.1\times10^{27}/\lambda_0^2(\text{nm})$ and here $\lambda_0=800\,$nm. Specifically, through a parametric experimental campaign performed at ``Laboratoire d'Optique Appliquée" (LOA), the charge-per-Joule metric serves as a straightforward method in gauging the efficiency of the different configurations of interest. By varying the laser energy, plasma density and gas nozzles, we study the conditions to produce highly divergent (i.e., $>100$s\,mrad), few-MeV electron beams with charges of $5-30\,$nC and a maximum laser-to-electron energy conversion efficiency around $14\,$\%. Through further investigation of the experimental configurations using Fourier-Bessel Particle-In-Cell (FBPIC)~\cite{fbpic} simulations, we identify three acceleration mechanisms: Ponderomotive Acceleration (PA), Wakefield Acceleration (WA) and Direct Laser Acceleration (DLA). Specifically, the numerical analysis underlines that, upon interaction with nitrogen, the driver pulse radially expels most electrons through its ponderomotive force. Consequently, our study challenges the wakefield's role in accelerating electrons within the configurations of interest, as commonly attributed.

\vspace{0.5cm}
%\vfill

\begin{figure*}[!htb]
\centering
\includegraphics[width=\textwidth]{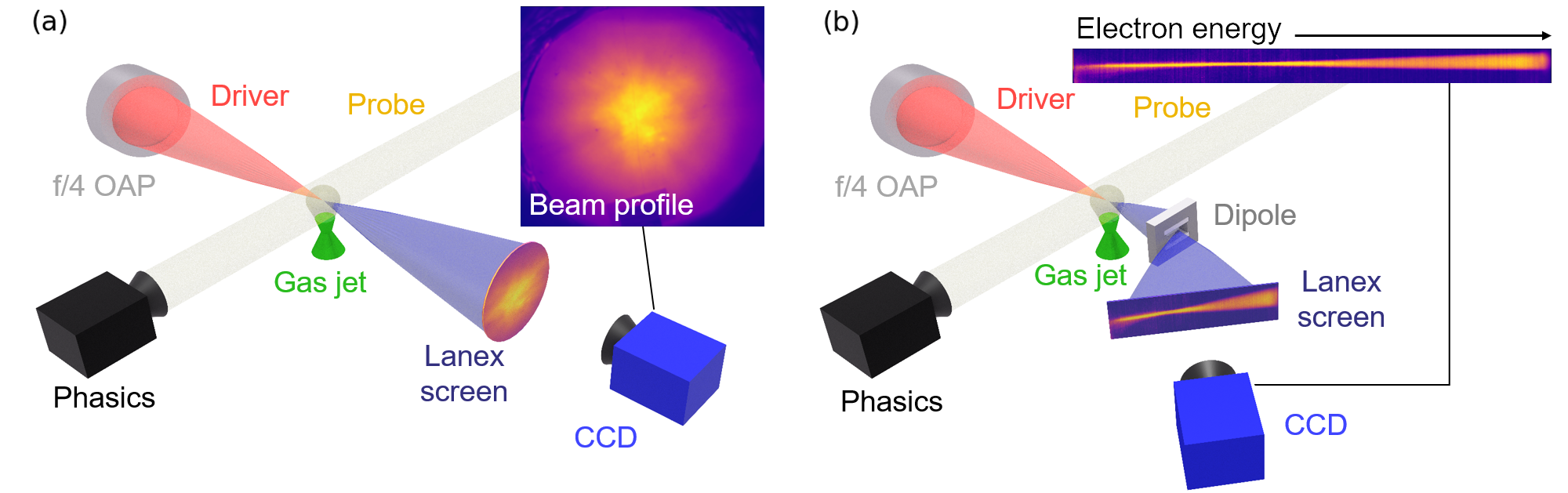}
\caption{Schematic representations of the experimental setup. The driver beam is focused using an f/4 off-axis parabola onto a supersonic gas jet. A transverse probe beam is employed to measure plasma density with a Phasics wavefront sensor. (a) A motorized beam profile monitor and (b) a motorized energy spectrometer are used for charge and energy measurements, respectively.}
\label{fig:schema_manip_fin}
\end{figure*}

%%%%%%%%%%%%%%%%%%%%%%%%%%%%%%%%%%%%%%%%%%%%%%%%%%%%%%%%%
\section{\label{sec:level2}Highly-Efficient Acceleration of Charged Electron Beams}
The experiment was performed using Salle Jaune's $60\,$TW Ti:Sapphire laser system able to produce linearly polarized pulses of $\lambda_0 = 813\,$nm central wavelength and a Full Width at Half Maximum (FWHM) duration of $30\,$fs, the driver beam in Figs.~\ref{fig:schema_manip_fin}(a) and (b). The laser pulse is focused using an f/4 Off-Axis Parabola (OAP) on the gas target, as shown in Fig.~\ref{fig:schema_manip_fin}(b), leading to a FWHM focal spot of $5\pm 0.3$\,$\mu$m. We estimate that the maximum laser energy on target is $1.7\pm0.2\,$J of which $\sim 57\,$\% is contained within the central spot, corresponding to the first minima ring. This corresponds to a peak laser intensity $I_0 \approx 1.5\times10^{20}\,$W\,cm$^{-2}$ and a maximum normalized vector potential $a_0 \approx 8.5$. We use pure nitrogen and two different cylindrically symmetric supersonic gas nozzles of $2\,$mm and $0.4\,$mm exit diameter. The latter is a fused silica nozzle obtained via hybrid 3D laser machining technique~\cite{Tomkus} produced at the Center for Physical Sciences and Technology in Lithuania. As depicted in Fig.~\ref{fig:schema_manip_fin}(b), using a Phasics wavefront sensor~\cite{JPrimot} and a probe laser pulse we are able to perform plasma density measurements.
\begin{figure*}[htbp]
    \centering
    \includegraphics[width=0.85\textwidth]{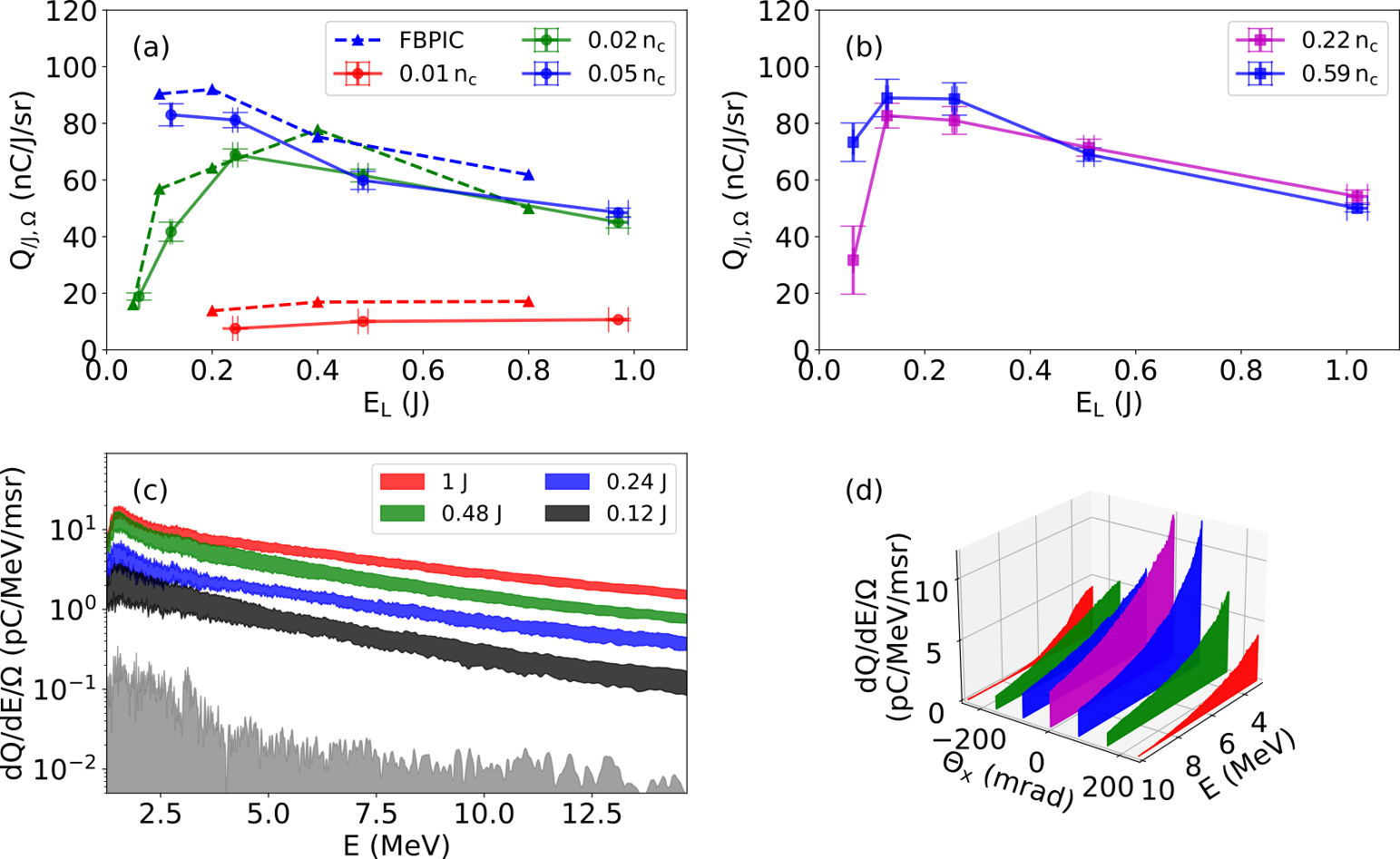}
    \caption{Experimental results. (a) Charge-per-Joule measured with the $0.4\,$mm nozzle in a $0.25\,$sr BPM solid angle. The horizontal error bar is the Root-Mean-Square Deviation (RMSD) on the laser energy calculated over one day. The vertical error bars represent the RMSD on ten shots on the charge-per-Joule. The dashed lines represent FBPIC simulation results. (b) Charge-per-Joule measured with the $2\,$mm nozzle in a $0.5\,$sr BPM solid angle. (c) Energy spectra obtained with the $2\,$mm nozzle at $n_e=0.29\,n_c$ for different laser energies. Each colored stripe is the RMSD over ten shots, while the gray surface represents the detection limit. (d) Energy spectra measured at different horizontal positions, expressed in terms of the horizontal angle $\Theta_x\in \left[-245,\,245\right]\,$mrad (i.e., perpendicular to the laser axis), in steps of $80\,$mrad. Here, we considered $n_e=0.29\,n_c$ and $E_L=1\,$J. Compared to the energy spectra in (c), these spectra are determined as the average over five shots.}
    \label{fig:res_fin}
\end{figure*}
This figure also illustrates the Beam Profile Monitor (BPM), allowing to perform shot-to-shot charge and beam divergence measurements. Specifically, the beam charge is determined using an absolutely calibrated tritium capsule~\cite{Kurz}. For each configuration, we calculate the average beam charge over ten consecutive shots, with the statistical error defined as the Root Mean Square Deviation (RMSD). The BPM consists of a motorized Lanex Regular Carestream screen with a diameter of $75\,$mm, positioned on the laser axis, and a 16-bit CCD camera to collect the electron beam within a given solid angle. To measure the energy spectra, instead, we remove the BPM and position a motorized electron spectrometer on the laser axis, as shown in Fig.~\ref{fig:schema_manip_fin}(b). Each spectrum is determined as the average over ten shots, with the corresponding RMSD. The spectrometer is composed of a magnetic dipole ($B_{max} = 0.44\,$T) with a $2\,$mm diameter pinhole at the entrance. The selected electrons are subsequently deviated on a Lanex screen calibrated from $1.2\,$MeV to $14.7\,$MeV.

%%%%%%%%%%%%%%%%%%%%%%%%%%%%%%%%%%%%%%%%%%%%%%%%%%%%%%%%%%%%%%%%%%
In Fig.~\ref{fig:res_fin}(a) the continuous curves represent the charge-per-Joule (Q$_{/J,\Omega}$) achieved with the $0.4\,$mm nozzle within a $\sim0.25$\,sr solid angle of the BPM, for three different plasma densities. Here, $E_L$ represents the estimated laser central spot energy. The PIC simulations (dashed lines) reproduce the charges obtained under these experimental conditions. Further details about this set of simulations can be found in the Supplementary Materials (Figs.~S1 and S2). In the experimental measurements, at $n_e = 0.01\,n_c$ we notice that the charge-per-Joule from $7.5\,$nC/J/sr at $E_L=0.24\,$J increases to around $10\,$nC/J/sr for $E_L=0.48\,$J. At higher plasma densities, instead, we observe a stronger dependency on the laser energy. Specifically, for $n_e=0.02\,n_c$ the charge-per-Joule increases with the laser energy, until reaching $E_L>0.24\,$J, where it reaches a maximum of $\sim69\,$nC/J/sr and subsequently starts to decrease. Similarly, the curve at $n_e=0.05\,n_c$ tends to decrease with the laser energy. The simulations show that this behavior is due to the limited BPM collecting angle: the dimension of the electron beams produced at these laser energies exceeds that of the BPM scintillating screen. Hence, some electrons are not co    llected by the diagnostics, causing a decrease in the charge-per-Joule. 

Furthermore, in Fig.~\ref{fig:res_fin}(a) we also notice that the curves at $n_e = 0.02\,n_c$ and $n_e = 0.05\,n_c$ superimpose, highlighting the presence of a charge-per-Joule saturation effect. However, the numerical study seems to confirm that the superposition of these curves is somehow magnified by the limited collecting angle. Indeed, even an increase in the plasma density yields more divergent electrons that fall out of the BPM measuring cone. With this nozzle, we estimate a maximum of $82\,$nC/J/sr at $E_L = 0.12\,$J (i.e., corresponding to a total charge of $2.5\,$nC), with a conversion efficiency $\eta = 6$\,\%. Here, we define the conversion efficiency as the ratio of the total electron energy to the total laser energy on target. 

In Fig.~\ref{fig:res_fin}(b), instead, we show the results obtained with the $2\,$mm nozzle at $n_e = 0.29\,n_c$ and $n_e = 0.59\,n_c$. In this figure, we move the BPM closer to the nozzle, allowing the collecting cone to increase to around $0.5\,$sr. In Fig.~\ref{fig:res_fin}(b) we notice very similar tendencies and values to what we have previously discussed. With this nozzle, we measure a maximum charge-per-Joule of $89\,$nC/J/sr with $n_e = 0.59\,n_c$ at $E_L = 0.12\,$J (i.e., $5.7\,$nC), corresponding to an energy conversion efficiency $\eta=14.4\,$\%. At full laser energy and at $n_e = 0.22\,n_c$, we reach $55\,$nC/J/sr, allowing to measure a maximum charge of $\sim 28$\,nC and an efficiency $\eta = 9.2\,$\%. The same efficiency is also estimated for the same laser energy at $n_e=0.59\,n_c$.

Fig.~\ref{fig:res_fin}(c) and (d) illustrate examples of electron energy spectra measured around the beam center. The curves of Fig.~\ref{fig:res_fin}(c) were obtained with the $2\,$mm nozzle at $n_e=0.29\,n_c$ for different laser energies. Thus, we observe that the shape of the energy spectrum and the beam's average energy remain constant across different laser energies ranging from $0.12\,$J to $1\,$J, with an average energy of approximately $5.25\pm0.12\,$MeV, calculated within the range $1.2\,$MeV and $14.7\,$MeV. We also measured the electron beam energy at different horizontal angles as shown in Fig.~\ref{fig:res_fin}(d). From this figure, it is possible to conclude that the most divergent electrons are also the least energetic. Indeed, the average energy drops from $\sim 6\,$MeV at the central position (i.e., $\Theta_x$ = 0) to $\sim4.5\,$MeV at the extremes (i.e., $\Theta_x$ = $\pm245\,$mrad), within the range $3-10\,$MeV.

\begin{figure}[!htb]
    \centering
    \includegraphics[width=0.5\textwidth]{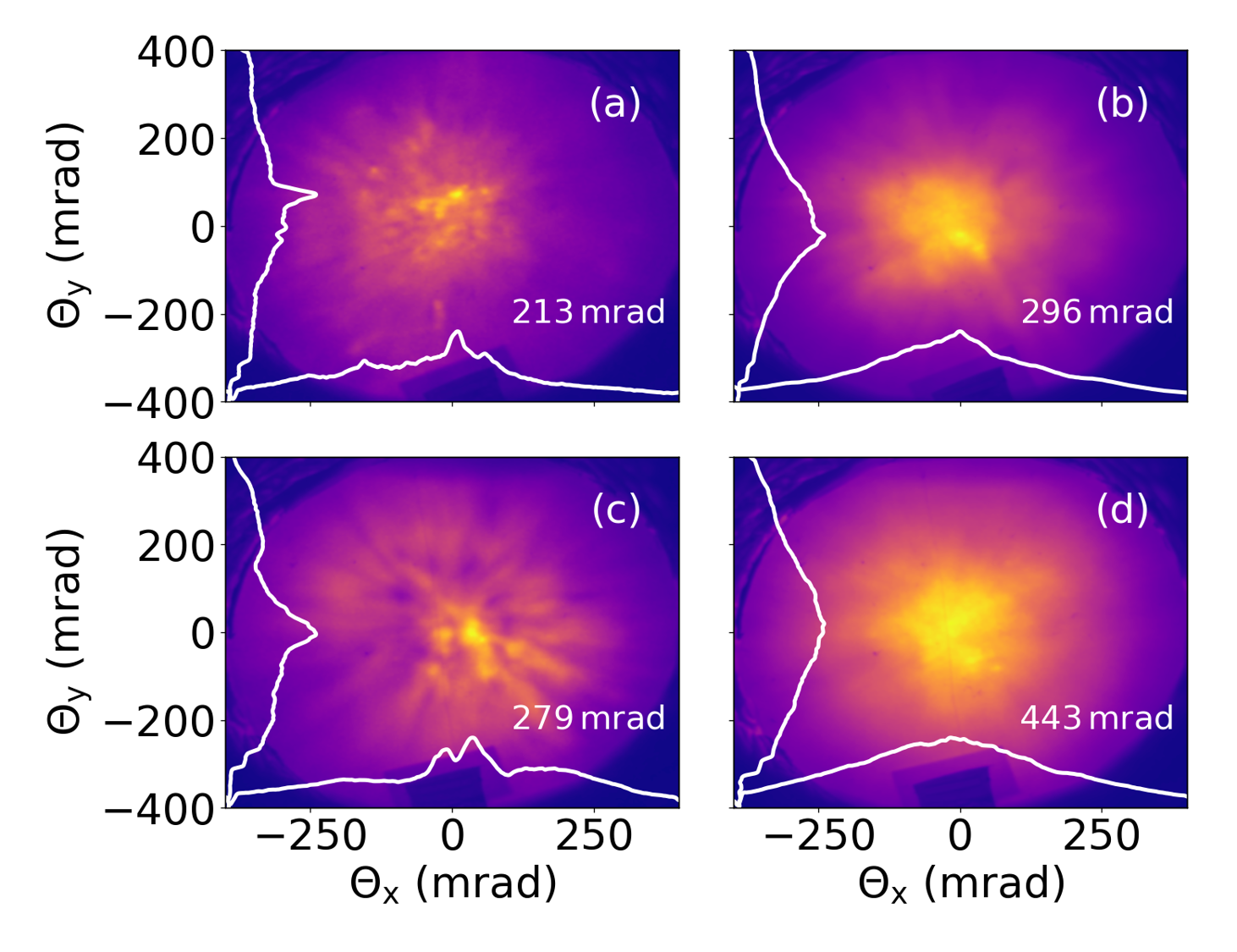}
    \caption{BPM single shot images obtained with the $2\,$mm nozzle for (a) $n_e = 0.29\,n_c$ and $E_L = 0.13$\,J (b) $n_e = 0.29\,n_c$ and $E_L = 1$\,J (c) $n_e = 0.59\,n_c$ and $E_L = 0.13$\,J (d) $n_e = 0.59\,n_c$ and $E_L = 1$\,J. The value on each image refers to the FWHM beam divergence.}
    \label{fig:beam_prof}
\end{figure}

In Fig.~\ref{fig:beam_prof}(a-d) we present some single-shot images of the BPM scintillating screen. Specifically, these images refer to the 2\,mm nozzle at $n_e = 0.29\,n_c$ and $n_e = 0.59\,n_c$ for two different laser energies. The white lines represent the transverse and longitudinal beam profiles passing through the maxima of each image. Coherent with our prior discussion, Figs.~\ref{fig:beam_prof}(a-d) prove that increasing either the plasma density or the laser energy leads to larger beam sizes. Indeed, from these figures, we estimate FWHM divergences between $200\,$mrad and $440\,$mrad, highlighting the influence of varying plasma density and laser energy on the beam divergence.
\vspace{1.8cm}

%%%%%%%%%%%%%%%%%%%%%%%%%%%%%%%%%%%%%%%%%%%%%%%%%%%%%%%%%
\section{\label{sec:level3} Numerical Analysis}
\subsection{Simulation setup}
In this section, we present a numerical study that allows us to interpret the experimental results. We first investigate the charge-per-Joule saturation effect and then discuss the different acceleration mechanisms.

The numerical study was performed employing the 3D code FBPIC, which uses a cylindrical grid with azimuthal decomposition. Concerning the simulation setup, we define a $30$\,$\mu$m density ramp, allowing the laser pulse to focus with limited energy losses at the beginning of a $1500$\,$\mu$m plateau (Fig.~S3 in Supplementary Materials). We assume the nitrogen gas is preionized up to N$^{+3}$, corresponding to the first three L-shell electrons, for numerical ease. The investigation considers four different plateau plasma densities between $n_e=0.02\,n_c$ and $n_e=0.18\,n_c$, corresponding to the full L-shell ionization (i.e., N$^{+5}$). The laser considered is a 30\,fs-Gaussian beam propagating along $z$ and polarized in the $x$ direction, with energies ($E_L$) ranging from $0.05\,$J to $1\,$J and a waist $w_0=3\,\mu$m. Regarding the numerical parameters, we employ a $(r,z)$ mesh with $\Delta z = \lambda_0 / 24$ and $\Delta r = 5\,\Delta z$, where $\lambda_0 = 800$\,nm is the laser wavelength. Finally, three azimuthal modes $(m=0-2)$ are considered, and the macroparticles per cell along $r,\,z$ and $\theta$ are set to 1, 1 and 4 respectively.

\subsection{Efficiency saturation effect}
\label{sec:ese}
\begin{figure}[!t]
    \centering
    \includegraphics[width=0.5\textwidth]{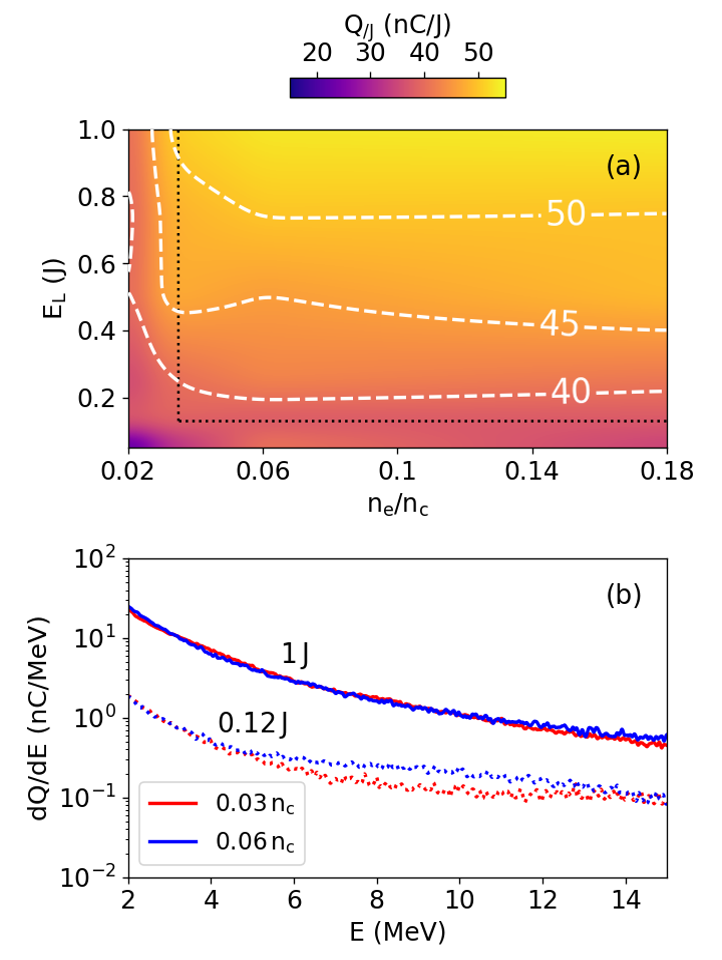}
    \caption{Charge-per-Joule for the different configurations of interest, considering electrons with energies $E>2$\,MeV. The colormap is the linear interpolation of the simulations, while the white dashed lines refer to charge-per-Joule isolines. (b) Examples of energy spectra. The continuous lines refer to the case at $E_L = 1$\,J for $n_e=0.03\,n_c$ and $n_e=0.06\,n_c$. The dotted curves refer to the case at $E_L=0.12$\,J at $n_e=0.03\,n_c$ and $n_e=0.06\,n_c$.}
    \label{fig:Q_sims}
\end{figure}

We now intend to explain the charge-per-Joule saturation effect observed experimentally and discussed in the Section~\ref{sec:level2}. Thus, in Fig.~\ref{fig:Q_sims} we present the charge-per-Joule (Q$_{/J}$) within a $4\pi\,$sr solid angle as a function of the laser energy and plasma density, derived from linearly interpolated numerical results. We only consider electrons with a minimum energy $E=2\,$MeV, which can be relevant for a number of low-energy applications such as industrial X-ray tomography. The white dashed lines in Fig.~\ref{fig:Q_sims}, instead, refer to charge-per-Joule isolines. From this figure, it is possible to notice that for $E_L\gtrsim0.12$\,J and $n_e\gtrsim0.03\,n_c$ the charge-per-Joule slowly increases from $40\,$nC/J to around $50\,$nC/J with the laser energy. This region is outlined by black dotted lines for visual reference. In accordance with the experimental findings discussed in Section~\ref{sec:level2}, we observe that within this region, increasing the plasma density at a fixed laser energy does not result in higher charges-per-Joule.

The numerical analysis underlines that this effect is due to the saturation of the conversion efficiency with respect to the plasma density. More precisely, the laser interacts with the plasma over a characteristic length approximately equal to half the pump depletion length, $L_{pd}\propto n^{-1}_e$~\cite{Lu2007}. Simultaneously, increasing the plasma density results in more electrons being accelerated per unit length. Therefore, despite the reduced interaction length at higher densities, the faster ionization ultimately leads to the same amount of charge being accelerated. This analysis is corroborated by the energy spectra in Fig.~\ref{fig:Q_sims}(b), where we consider electrons with energies in the range $2\,\text{MeV}<E<15\,\text{MeV}$, comprising over $80\,$\% of the charge above $2\,$MeV. Indeed, we notice that the curves at $E_L=1\,$J for $n_e=0.03\,n_c$ and $n_e=0.06\,n_c$ overlap, presenting an average energy of $4.7\,$MeV. For these configurations, we estimate the efficiency to be $\eta\sim21\,$\%. A similar behavior is also observed at $E_L=0.12\,$J, where we estimate the average energy to be $5.7\,$MeV and the conversion efficiency is $\eta\sim18\,$\% for both densities. In other words, regardless of variations in plasma density, the same amount of laser energy ionizes and accelerates the same number of electrons to the same average energy.

\subsection{Electron dynamics and acceleration mechanisms}
\begin{figure*}[!htb]
    \centering
    \includegraphics[width=\textwidth]{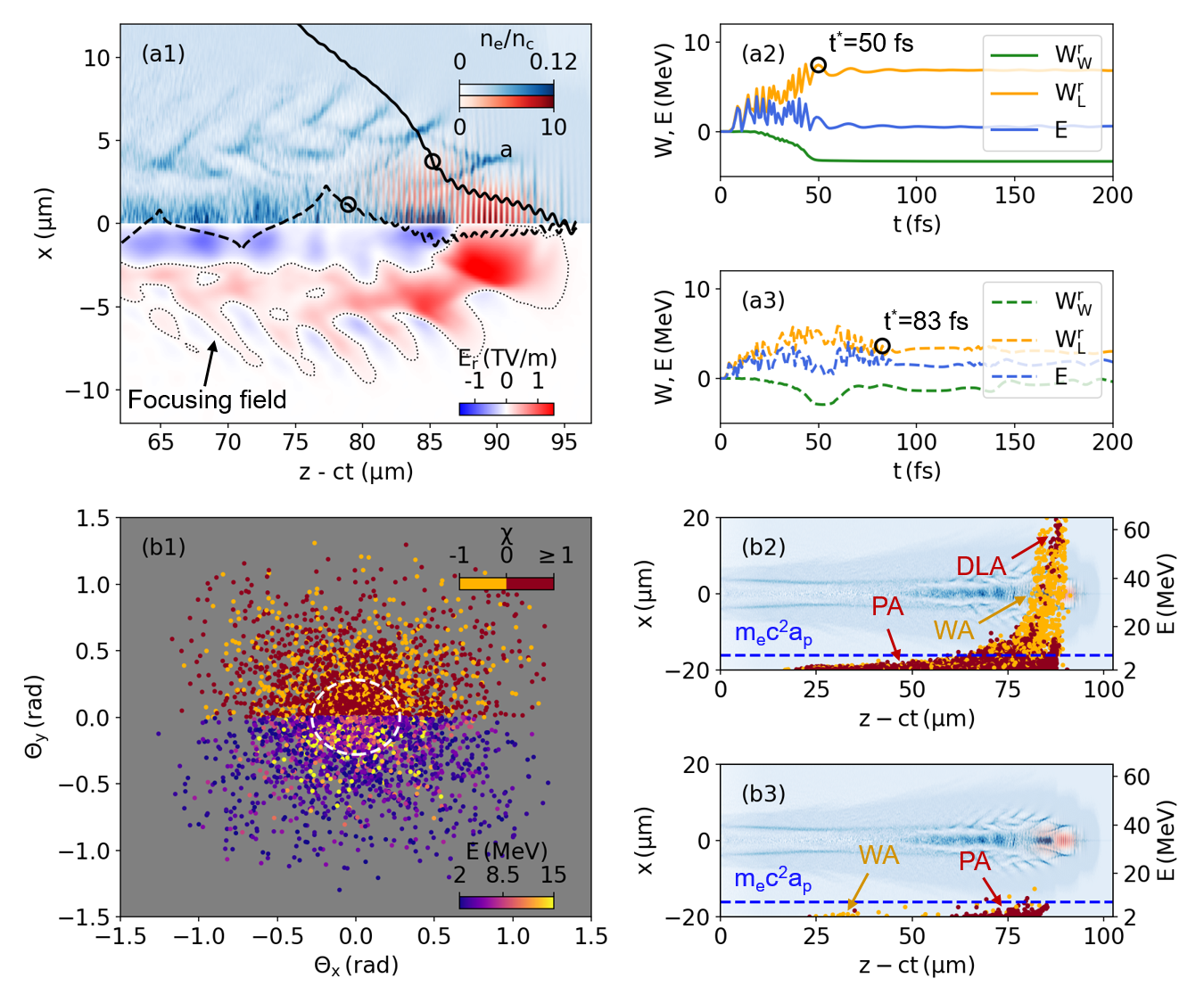}
    \caption{FBPIC results with $n_e=0.03\,n_c$ and $E_L=1\,$J. (a1) In the top half, the plasma density is shown in blue and the laser pulse in red. The bottom half presents the radial wakefield. The black curves represent two K-shell electrons undergoing ponderomotive acceleration. The circles on each trajectory indicate the time when the electron exits the laser field. (a2-3) Radial Wakefield ($W^r_W$) and laser work ($W^r_L$), along with electron kinetic energy for the continuous and dashed trajectory in (a1). (b1) Electron angular distribution. In the top half, the color refers to the laser-to-wakefield work ratio $\chi$ (see text), while in the bottom half, it represents the energy. $(z-ct,\,E)$ phase space for (b2) K-shell electrons and (b3) L-shell electrons. For visual reference, we show the plasma density and laser intensity.}
    \label{fig:K_shell_work}
\end{figure*}

Before discussing the acceleration mechanisms in detail, we intend to briefly describe the plasma structures that develop during the interaction. Thus, Fig.~\ref{fig:K_shell_work}(a1) depicts the plasma density (top half) and corresponding radial plasma field (bottom half) obtained with $n_e=0.03\,n_c$ and $E_L=1\,$J. In this figure, we notice the presence of a long and rapidly-changing channel-like structure, with a characteristic length $L\gg\lambda_p$, where $\lambda_p=4.7\,\mu$m is the plasma wavelength. This massive sheath is formed by nitrogen L-shell electrons and it is filled with K-shell electrons, continuously accumulating and flowing through the channel.
In the bottom half of Fig.~\ref{fig:K_shell_work}(a1), we highlight that the radial focusing field generated by the L-shell electron sheath confines K-shell electrons close to the laser axis. Consequently, this dense concentration of particles on axis shields the longitudinal wakefield and hinders the formation of ion cavities. Nonetheless, inside of this structure, we can still find rapidly changing density modulations, which can generate a longitudinal wakefield. As we will discuss in the following, these modulations can contribute to the electron acceleration process. Close to the laser intensity peak (i.e., $z-ct\approx90\,\mu$m), instead, the ponderomotive force allows for effective charge separation and the formation of a cavity in front of the channel structure.

In order to understand how electrons gain energy, we calculated the work performed by the laser and plasma electric fields. This task was carried out employing the numerical tool FBPIC-Electric Work Profiler (FBPIC-EWP)~\cite{Martelli2024}. This code allows to estimate the work exerted on FBPIC-tracked electrons by both the laser and plasma electric fields, exploiting FBPIC modal decomposition. Hence, if we consider a single electron, in the time interval $[0, t]$ an electric field $\mathbf{E}_{W,L}$ performs the work
\begin{equation}
    W_{W,L}(t) = -e\int^{t}_{0} \mathbf{E}_{W,L} \cdot \mathbf{v}\,dt',
\end{equation}
where the subscripts $W$ and $L$ denote the wakefield and laser field contribution respectively, $\mathbf{v}$ is the electron velocity and $e$ is its charge.

Fig.~\ref{fig:K_shell_work}(a1) illustrates two examples of K-shell electron trajectories (black curves) on the $(z-ct,\,x)$ plane experiencing ponderomotive acceleration. L-shell electrons exhibit a similar dynamic as will be discussed in the following and in the Supplementary Materials (Fig.~S6). Considering the case denoted by the continuous trajectory in Fig.~\ref{fig:K_shell_work}(a1), we estimate that the particle crosses $N_{oc} = 12$ laser optical cycles before being expelled at the instant $t^*=50\,$fs, denoted by a circle in Fig.~\ref{fig:K_shell_work}(a1). Here, $t=0$ corresponds to the electron ionization time. Similar to what is observed in other acceleration regimes~\cite{Thevenet2016}, this sort of dynamic is typical of ponderomotive electrons: they slip through several optical cycles experiencing low energy gains, as shown in Fig.~\ref{fig:K_shell_work}(a2). Here, we plot the electron kinetic energy and, for simplicity, the work done by the plasma and the laser along the radial direction, defined as $W^r_{W,L}=W^x_{W,L} + W^y_{W,L}$. We estimate that the laser provides $6.8\,$MeV in radial push. Simultaneously, the electron loses $3.2\,$MeV crossing regions of space where the focusing radial wakefield opposes the ponderomotive push (i.e., at $z-ct\approx85\,\mu$m). Concerning the longitudinal dynamics, instead, the numerical analysis underlines that at $t=22\,$fs (i.e., $z-ct\approx93\,\mu$m) the electron crosses the wakefield decelerating region in the front cavity, causing a $1\,$MeV loss, while also the laser performs a negative work around $2.1\,$MeV. Finally, the electron leaves the laser with an energy $E=0.5\,$MeV.

The particle denoted with the dashed trajectory in Fig.~\ref{fig:K_shell_work}(a1) experiences a somewhat similar acceleration process. It slips through the laser field oscillating over $N_{oc} = 21$ optical cycles, before leaving the laser field at $t^*=83\,$fs, as denoted by the circle on the electron trajectory. From Fig.~\ref{fig:K_shell_work}(a3) we deduce that in this case, the laser also provides most of the energy, with a radial push of $2.7\,$MeV, while the wakefield exerts a pull of $1.1\,$MeV. Longitudinally, the particle receives $0.7\,$MeV from the plasma field, since it experiences its accelerating phase at $t=50\,$fs (i.e., $z-ct\approx86\,\mu$m), while the laser performs a negative work of $0.8\,$MeV. The electron leaves the laser pulse with an energy of $E=1.5\,$MeV. Moreover, for $t>83\,$fs in Fig.~\ref{fig:K_shell_work}(a1) we note that this particle remains closer to the laser axis and is radially trapped in the channel-like structure discussed above. Here, it performs radial oscillations until leaving the channel without a significant difference in energy.

Having discussed the ponderomotive acceleration through two examples, we now present a statistically relevant study distinguishing between the number of electrons accelerated by the laser or the wakefield. Moreover, we provide an overview of other acceleration mechanisms we identified, and more details can be found in the Supplementary Materials (Figs.~S4 and S5). Therefore, for each tracked electron at the instant $t$, we now define the laser-to-plasma work ratio as
\begin{equation}
    \chi(t) = \left|\frac{W_{L}(t)}{W_{W}(t)}\right| - 1,
\end{equation}
allowing to compare the laser and plasma field contributions. A positive $\chi$ value clearly indicates the laser's central role in driving the electron acceleration process. For coherence with Section~\ref{sec:ese}, in Fig.~\ref{fig:K_shell_work}(b1) we plot the beam angular distribution for both K- and L-shell electrons with $E>2\,$MeV at the last iteration. Each electron in the top half of this figure is colored based on its $\chi$ value. Thus, we define two criteria for $\chi$ and we identify the corresponding electron populations. In the range $-1\leq\chi<0$ we find yellow electrons that gain most of their energy from wakefield acceleration.
\begin{figure}[htbp]
    \centering
    \includegraphics[width=0.5\textwidth]{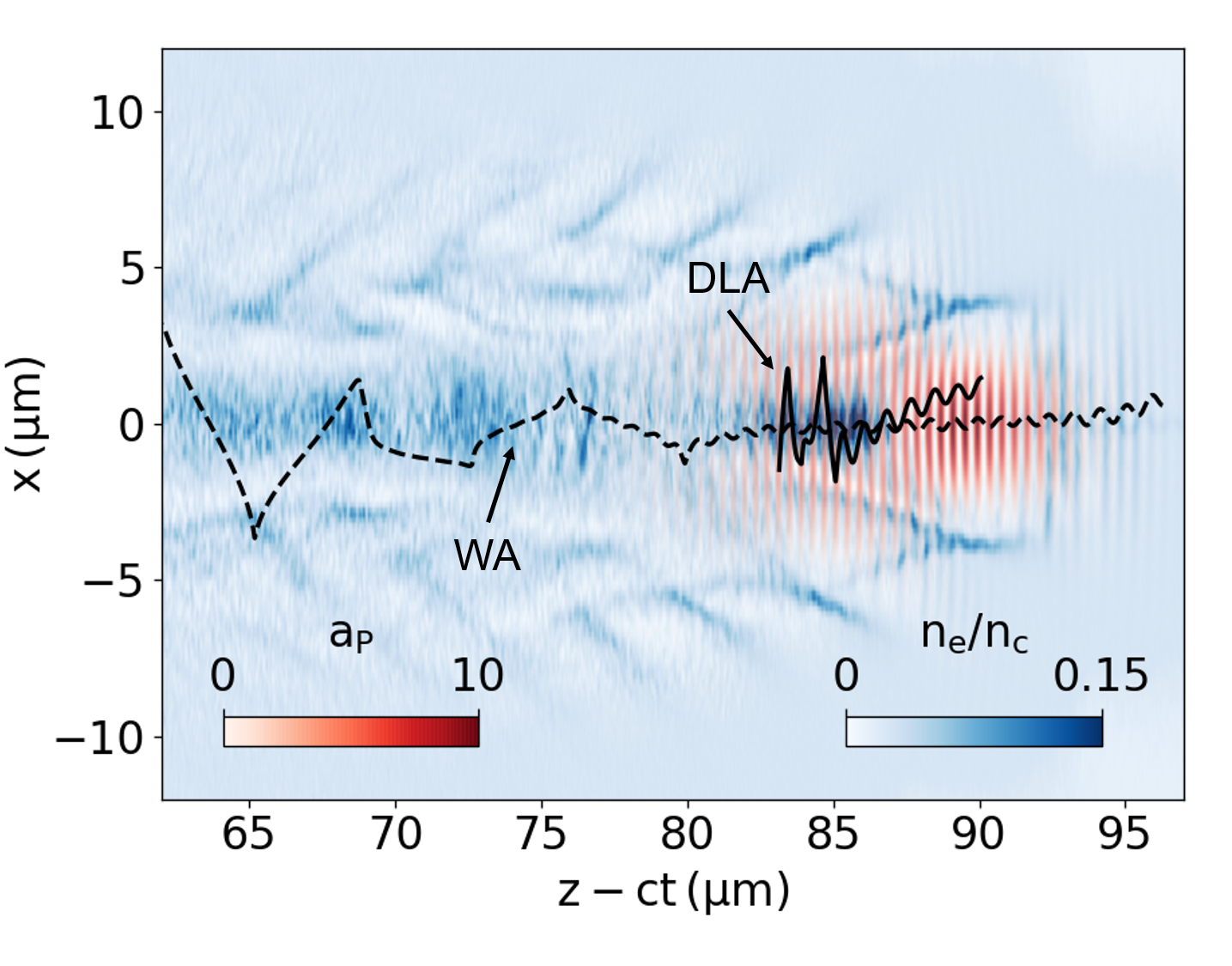}
    \caption{Trajectory examples of electrons undergoing (dashed curve) wakefield acceleration and (continuous curve) direct laser acceleration. More details are provided in Fig.~S4 and S5 in the Supplementary Materials.}
    \label{fig:wake_dla}
\end{figure}
The red particles, instead, have $\chi\geq0$ and are primarily accelerated by the laser. In this population, we find ponderomotive electrons, displaying dynamics similar to the two examples in Fig.~\ref{fig:K_shell_work}(a1). As discussed below, some of these electrons can also experience direct laser acceleration~\cite{Pukhov1999, Pukhov2003, Arefiev}. We estimate that around $70\,$\% of the particles exhibit positive values of $\chi$, proving the laser as the primary driver in particle acceleration. In the bottom half of Fig.~\ref{fig:K_shell_work}(b1), instead, the color represents the energy of each particle. The average energy for $E>2\,$MeV is around $9\,$MeV for both laser- and wakefield-accelerated electrons, and we notice a concentration of higher energies closer to the laser axis, which is consistent with the experimental observation. For instance, if we consider a solid angle of $0.25\,$sr (white circle in Fig.~\ref{fig:K_shell_work}(b1)), the average energy increases to $14\,$MeV. 

Fig.~\ref{fig:K_shell_work}(b2) depicts the $(z-ct,\,E)$ phase space of laser- and wakefield-accelerated K-shell electrons at the final iteration of the simulation. This figure highlights the different behaviors of laser-accelerated electrons. As previously mentioned, in this population we find particles undergoing ponderomotive acceleration. We estimate that around $85\,$\% of laser-accelerated K-shell electrons gain up to $m_ec^2a_p\approx8\,$MeV through the ponderomotive push~\cite{Macchi}. Here, $a_p\approx16$ is the maximum laser normalized vector potential in plasma. Instead, the remaining $15\,$\% experiences direct laser acceleration. Fig.~\ref{fig:wake_dla} presents an example of electron trajectory undergoing DLA (continuous curve). These particles are trapped in the ion cavity in front of the channel structure via ionization injection. Once trapped, they overlap with the driver pulse and perform oscillations along the laser polarization direction. Subsequently, these oscillations can lead to a gain in longitudinal momentum via the $e\mathbf{v}\times\mathbf{B}$ term of the laser. The longitudinal wakefield reduces the dephasing between the electrons and the laser, ultimately allowing these particles to reach $10\text{s}-100\text{s}$\,MeV~\cite{Shaw2014, Shaw2016}. In the example of Fig.~\ref{fig:wake_dla}, the electron reaches a maximum energy of $85\,$MeV, with $79\,$MeV attributed to the laser.

In Fig.~\ref{fig:K_shell_work}(b2) we can also observe wakefield-accelerated K-shell electrons (yellow dots). Specifically, the numerical analysis highlights the presence of ``traditional" wakefield acceleration, where the particles are trapped in phase with the wakefield within the front ion cavity long enough to reach $10\text{s}-100\text{s}\,$MeV. Additionally, we recognize wakefield accelerated particles displaying a behavior somewhat similar to the dashed trajectory of Fig.~\ref{fig:K_shell_work}(a1). Once ionized, they initially receive most of their energy from the laser ponderomotive push and they subsequently slip into the channel, where they radially oscillate until escaping the structure with few-MeVs in energy. However, once inside the channel, they undergo acceleration in the longitudinal wakefield phase, induced by the plasma density modulations mentioned earlier. The numerical analysis shows that this plasma field contribution exceeds the laser initial push and, in this sense, they undergo wakefield acceleration. Nonetheless, unlike traditional wakefield acceleration, where particles are trapped in phase with the wakefield within the front ion cavity, these electrons continuously slip through the channel. In Fig.~\ref{fig:wake_dla} we provide an example of this sort of non-traditional wakefield acceleration (dashed curve). At the end of the simulation, the particle has gained $4.3\,$MeV in energy, of which $\sim3\,$MeV are provided by the wakefield.

Analogously, Fig.~\ref{fig:K_shell_work}(b3) displays the $(z-ct,\,E)$ phase space of laser- and plasma-accelerated L-shell electrons at the final iteration of the simulation. We estimate that around $83\,$\% of L-shell electrons with $E>2\,$MeV undergo PA, while the remaining fraction is mainly accelerated via WA while momentarily crossing the longitudinal wakefield accelerating phase.

\section{\label{sec:level4} Conclusion and Outlook}
In this paper, we studied a regime allowing the increase of the average current of laser-plasma accelerators with energies at few-MeVs, through the enhancement of the beam charge. With an extensive experimental campaign, we were able to produce charges of $5-30\,$nC, with average energies around $5\,$MeV. Notably, employing a $0.12\,$J laser pulse, we achieve a charge of $5.3\,$nC with a conversion efficiency of $14.4\,$\%, one of the highest recorded to date. Improving the laser focal spot quality could lead to efficiencies exceeding $20\,$\%, as shown by the numerical study. These results are promising for future LPAs, paving the way for unprecedented average currents. Novel J-class lasers with $100\,$W in average power are emerging~\cite{Pellegrina2022, Kiani2023} and they would allow to exceed the $\mu$A-level. Considering, for instance, the configuration discussed above (i.e., $5.7\,$nC at $0.12\,$J), we can easily estimate that the maximum achievable average current is $\sim5\mu$A, assuming $100\,$W in laser average power. With such capabilities, this electron source emerges as a promising candidate for various applications needing few-MeV electrons, including X-ray tomography~\cite{Svendsen2018, Cole2018} and irradiation studies.

In configurations similar to those studied here, electron trapping via ionization injection followed by wakefield acceleration is typically considered the dominant acceleration process. However, in this paper, we were able to show that most particles are not injected in plasma cavities, as also discussed in other works present in the literature~\cite{Yang2017, Behm2019}. More specifically, we proved that the electrons primarily gain energy through ponderomotive acceleration, establishing the laser pulse as the main driver in the energy gain mechanism.

\begin{acknowledgments}
This project has received funding from the European Union’s Horizon 2020 research and innovation program under grant agreement n°101020100.
\end{acknowledgments}

\bibliographystyle{apsrev4-2}
\bibliography{apssamp}

\end{document}